\documentclass[twocolumn]{aastex63}
\usepackage{graphicx}

\newcommand\planet{HD\,106315\,c}

\newcommand{\project}[1]{\textsl{#1}}   

\newcommand{\JWST}{\project{JWST}}        \newcommand{\HST}{\project{HST}}          \newcommand{\Spitzer}{\project{Spitzer}}   \newcommand{\Ktwo}{\project{K2}}

\shorttitle{Water in the atmosphere of \planet?}
\shortauthors{Kreidberg et al.}

\begin{document}

\title{Tentative Evidence for Water Vapor in the Atmosphere of the Neptune-Size Exoplanet \planet}

\correspondingauthor{Laura Kreidberg}
\email{kreidberg@mpia.de}

\author[0000-0003-0514-1147]{Laura Kreidberg}
\affiliation{Max-Planck-Institut f\"ur Astronomie, K\"onigstuhl 17, 69117 Heidelberg, Germany}
\affiliation{Center for Astrophysics $|$ Harvard \& Smithsonian, 60 Garden Street, Cambridge, MA, 02138, USA}

\author[0000-0003-4096-7067]{Paul Molli\`ere}
\affiliation{Max-Planck-Institut f\"ur Astronomie, K\"onigstuhl 17, 69117 Heidelberg, Germany}

\author{Ian J.M. Crossfield}
\affiliation{The University of Kansas, Department of Physics and Astronomy, Malott Room 1082, 1251 Wescoe Hall Drive, Lawrence, KS, 66045, USA}

\author[0000-0002-5113-8558]{Daniel P. Thorngren}
\affiliation{ Institute for Research on Exoplanets, Universit\'e de Montr\'eal, Montr\'eal, Quebec, H3T 1J4, Canada}

\author[0000-0003-3800-7518]{Yui Kawashima}
\affil{SRON Netherlands Institute for Space Research, Sorbonnelaan 2, 3584 CA Utrecht, The Netherlands}

\author[0000-0002-4404-0456]{Caroline V. Morley}
    \affiliation{Department of Astronomy, University of Texas at Austin, Austin, TX 78712, USA}
    
\author{Bj\"orn Benneke}
\affiliation{ Institute for Research on Exoplanets, Universit\'e de Montr\'eal, Montr\'eal, Quebec, H3T 1J4, Canada}

\author[0000-0001-5442-1300]{Thomas Mikal-Evans}
\affiliation{Kavli Institute for Astrophysics and Space Research, Massachusetts Institute of Technology, 77 Massachusetts Avenue, 37-241, Cambridge, MA 02139, USA}

\author[0000-0001-6298-412X]{David Berardo}
\affil{Kavli Institute for Astrophysics and Space Research, Massachusetts Institute of Technology, 77 Massachusetts Avenue, 37-241, Cambridge, MA 02139, USA}

\author{Molly Kosiarek}
\affiliation{Department of Astronomy \& Astrophysics, University of California,  Santa Cruz, CA 95064, USA}

\author[0000-0002-8990-2101]{Varoujan Gorjian}
\affiliation{MS 169-506, Jet Propulsion Laboratory, California Institute of Technology, 4800 Oak Grove Drive, Pasadena, CA 91109, USA}

\author[0000-0002-5741-3047]{David R. Ciardi}
\affiliation{Caltech/IPAC-NExScI, M/S 100-22, 770 S Wilson Ave, Pasadena, CA 91106 USA}

\author[0000-0002-8035-4778]{Jessie L. Christiansen}
\affiliation{Caltech/IPAC-NExScI, M/S 100-22, 770 S Wilson Ave, Pasadena, CA 91106 USA}

\author[0000-0003-2313-467X]{Diana~Dragomir}
\affiliation{Department of Physics and Astronomy, University of New Mexico, 1919 Lomas Blvd NE, Albuquerque, NM 87131, USA}

\author[0000-0001-8189-0233]{Courtney D. Dressing}
\affiliation{Department of Astronomy, 501 Campbell Hall, University of California at Berkeley, Berkeley, CA 94720, USA} 

\author[0000-0002-9843-4354]{Jonathan J. Fortney}
\affiliation{Department of Astronomy \& Astrophysics, University of California,  Santa Cruz, CA 
95064, USA}

\author[0000-0003-3504-5316]{Benjamin J. Fulton}
\affiliation{NASA Exoplanet Science Institute / Caltech-IPAC, Pasadena, CA 91125, USA}

\author[0000-0002-8963-8056]{Thomas P. Greene}
\affiliation{Space Science and Astrobiology Division, NASA Ames Research Center, MS 245-6, Moffett Field, CA 94035, USA}

\author[0000-0003-3702-0382]{Kevin K. Hardegree-Ullman}
\affiliation{Caltech/IPAC-NExScI, M/S 100-22, 770 S Wilson Ave, Pasadena, CA 91106 USA}

\author[0000-0001-8638-0320]{Andrew W. Howard}
\affiliation{California Institute of Technology, Pasadena, CA 91125, USA}

\author[0000-0002-2532-2853]{Steve~B.~Howell}
\affil{NASA Ames Research Center, Moffett Field, CA 94035, USA}

\author[0000-0002-0531-1073]{Howard Isaacson}
\affiliation{501 Campbell Hall, University of California at Berkeley, Berkeley, CA 94720, USA} \affiliation{Centre for Astrophysics, University of Southern Queensland, Toowoomba, QLD, Australia}

\author[0000-0002-2413-5976]{Jessica E. Krick}
\affiliation{Caltech/IPAC, M/S 330-6, 1200 E. California Blvd, Pasadena, CA 91125 USA}

\author[0000-0002-4881-3620]{John~H.~Livingston}
\affiliation{Department of Astronomy, University of Tokyo, 7-3-1 Hongo, Bunkyo-ku, Tokyo 113-0033, Japan}
 
\author[0000-0003-3667-8633]{Joshua D. Lothringer}
\affiliation{Department of Physics and Astronomy, Johns Hopkins University, Baltimore, MD 21218, USA}

\author[0000-0001-9414-3851]{Farisa Y. Morales}
\affiliation{MS 169-214, Jet Propulsion Laboratory, California Institute of Technology, 4800 Oak Grove Drive, Pasadena, CA 91109, USA}

\author[0000-0003-0967-2893]{Erik A Petigura}
\affiliation{Department of Physics \& Astronomy, University of California Los Angeles, Los Angeles, CA 90095, USA}

\author[0000-0001-8812-0565]{Joseph E. Rodriguez} 
\affiliation{Center for Astrophysics $|$ Harvard \& Smithsonian, 60 Garden Street, Cambridge, MA, 02138, USA}

\author[0000-0001-5347-7062]{Joshua E. Schlieder}
\affiliation{Exoplanets and Stellar Astrophysics Laboratory, Code 667, NASA Goddard Space Flight Center, Greenbelt, MD 20771, USA}

\author[0000-0002-3725-3058]{Lauren M. Weiss}
\affiliation{Institute for Astronomy, University of Hawai`i, 2680 Woodlawn Drive, Honolulu, HI 96822, USA}

\begin{abstract}
We present a transmission spectrum for the Neptune-sized exoplanet \planet\ from optical to infrared wavelengths based on transit observations from the Hubble Space Telescope/Wide Field Camera~3, \Ktwo, and \Spitzer.  The spectrum shows tentative evidence for a water absorption feature in the $1.1 - 1.7\mu$m wavelength range with a small amplitude of 30 ppm (corresponding to just $0.8 \pm 0.04$ atmospheric scale heights). Based on an atmospheric retrieval analysis, the presence of water vapor is tentatively favored with a Bayes factor of 1.7 - 2.6 (depending on prior assumptions). The spectrum is most consistent with either enhanced metallicity, high altitude condensates, or both.  Cloud-free solar composition atmospheres are ruled out at $>5\sigma$ confidence. We compare the spectrum to grids of cloudy and hazy forward models and find that the spectrum is fit well by models with moderate cloud lofting or haze formation efficiency, over a wide range of metallicities ($1 - 100\times$ solar). We combine the constraints on the envelope composition  with an interior structure model and estimate that the core mass fraction is $\gtrsim0.3$.  With a bulk composition reminiscent of that of Neptune and an orbital distance of 0.15 AU, \planet\ hints that planets may form out of broadly similar material and arrive at vastly different orbits later in their evolution.
\end{abstract}


\section{Introduction}
The origins of Uranus and Neptune remain mysterious. Based on current data, it is not known if they formed by core accretion or gravitational instability, whether they grew in their current locations or underwent migration, or how long it took them to form \citep[][and references therein]{atreya20}. One of the challenges in modeling these planets' origin is that their bulk composition is poorly constrained. Uranus and Neptune are so cold that many of the dominant carbon, nitrogen, and oxygen-bearing molecules have condensed out of the observable atmosphere, leaving only methane accessible by remote observation \citep{helled20}. There are calls for a space mission to explore one of the ice giants \emph{in situ} and measure their atmospheric abundances directly with a probe; however, such a mission is over a decade away \citep{simon20}.

Meanwhile the search for extrasolar planets has revealed an abundance of Neptune-size worlds \citep[e.g.][]{coughlin16}. Many of these have short orbital periods and correspondingly high equilibrium temperatures (up to 2000~K), meaning that major volatile species are expected to be in the gas phase in the observable part of the atmosphere \citep{moses13}. Atmosphere characterization of these hotter exo-Neptunes provides an opportunity to determine their chemical compositions, well in advance of \emph{in situ} measurements of the Solar System ice giants. 

Precise near-infrared transmission spectra are available for fewer than a dozen exoplanets in the Neptune-mass range, $10 - 40\,M_\oplus$ \citep{crossfield17, kreidberg18b, spake18, mansfield18, benneke19a, benneke19,roberts20, guo20, chachan19}. Planets of this size are expected to have modest H/He envelopes ($\gtrsim1$\% by mass), with a diversity of atmospheric metal enrichment \citep[e.g.][]{fortney13, wolfgang15}. The transmission spectra measured to date have a wide range of properties that match the diversity expected from theoretical models. Some planets appear to have very high metallicity envelopes \citep[e.g. the $\sim1000\times$ solar composition inferred for GJ~436b;][]{morley17}. Others have lower metallicity, more akin to Jupiter's  $<10\times$ solar composition \citep[HAT-P-26b;][]{wakeford17}. The planets also have a wide range of cloud and haze properties, from cloud-free to very high altitude condensates, which complicate the interpretation of the measured spectra \citep{kreidberg14a, crossfield17}. To fully explore the diversity of the exo-Neptune population and identify cloud-free atmospheres,  a larger sample size is needed, which is the goal of the ongoing large \HST\ program GO 15333 (PIs I. Crossfield and L. Kreidberg). In total this program will obtain transmission spectra for five additional Neptune-size exoplanets, including the subject of this work, \planet.

 First observed by the \Ktwo\ mission \citep{crossfield17b, rodriguez17}, \planet\ has a radius of $4.0\pm0.4\,R_\oplus$ and a mass of $15.2\pm3.7\,R_\oplus$ \citep{barros17}\footnote{We note that the planet mass and radius are being revised in a companion paper. We use the updated values in our analysis and will add them to the arXiv version of this paper upon submission of Kosiarek et al., in prep.}. The planet has a 21.06 day orbit around its F5-type host star, and an equilibrium temperature of $870\pm20$ K (assuming full heat redistribution and zero Bond albedo). Thanks to the bright host (H magnitude = 8.0), \planet\ is one of the most accessible candidates for atmosphere characterization with transmission spectroscopy, with a Transmission Spectroscopy Metric equal to 119 \citep[this metric is a proxy for the expected signal-to-noise of the transmission spectrum;][]{kempton18}. Compared to other exo-Neptunes with precise spectra, \planet\ has a longer period and a more massive host star \citep{crossfield17}. It is also part of a multi-planet system, with an interior $2.6\,R_\oplus$ planet orbiting the star every 9.6 days.  

\section{Observations}\label{sec:observations}
We observed four transits of \planet\ with the Wide Field Camera 3 instrument on the \emph{Hubble Space Telescope} (\HST/WFC3) as part of Program GO 15333 (Co-PIs: I. Crossfield and L. Kreidberg). The dates of the observations were 3 December 2018, 21-22 December 2018, 2 February 2019, and 21 November 2019. There was also an observation on 15 June 2018 that failed due to lost guiding.  Each transit observation consists of time series exposures over six continuous \HST\ orbits. The first exposure of each orbit was a direct image with the F126N filter. Subsequent exposures used the G141 grism, which covers the wavelength range from $1.1 - 1.7\mu$m. The exposures used the \texttt{SPARS25}, \texttt{NSAMP} = 8 readout mode which has a total integration time of 138.4 s. The observations used spatial scanning mode, which spreads the light in the cross-dispersion direction during the exposure, enabling longer integration times before saturation \citep{deming13}. The scan rate was 0.213"/sec, yielding a total scan height of 31" (238 pixels). We observed 14 exposures per orbit, for an observing efficiency of 73\%.

A single transit was also observed by the \emph{Spitzer} Space Telescope \citep{werner04, fazio04} with the IRAC2 4.5 $\mu$m photometric channel on 2017 April 19 - 20 as part of Program 13052 (PI: M. Werner). The observations used PCRS peak-up mode, which positions the target precisely on a pixel with minimal sensitivity variations\footnote{\url{https://irsa.ipac.caltech.edu/data/SPITZER/docs/irac/pcrs_obs.shtml}}. The observation began with a 30-minute stare to allow the spacecraft to thermally settle, followed by 32168 s (8.9 hours) of science data with an exposure time of 0.4 seconds. Two transits were also observed by \Ktwo\ \citep[previously described in][]{crossfield17,rodriguez17}.

\section{Data Reduction and Analysis}
\subsection{\HST/WFC3}
\label{sec:wfc3}
\begin{figure*}
\begin{center}
\includegraphics[scale = 0.5]{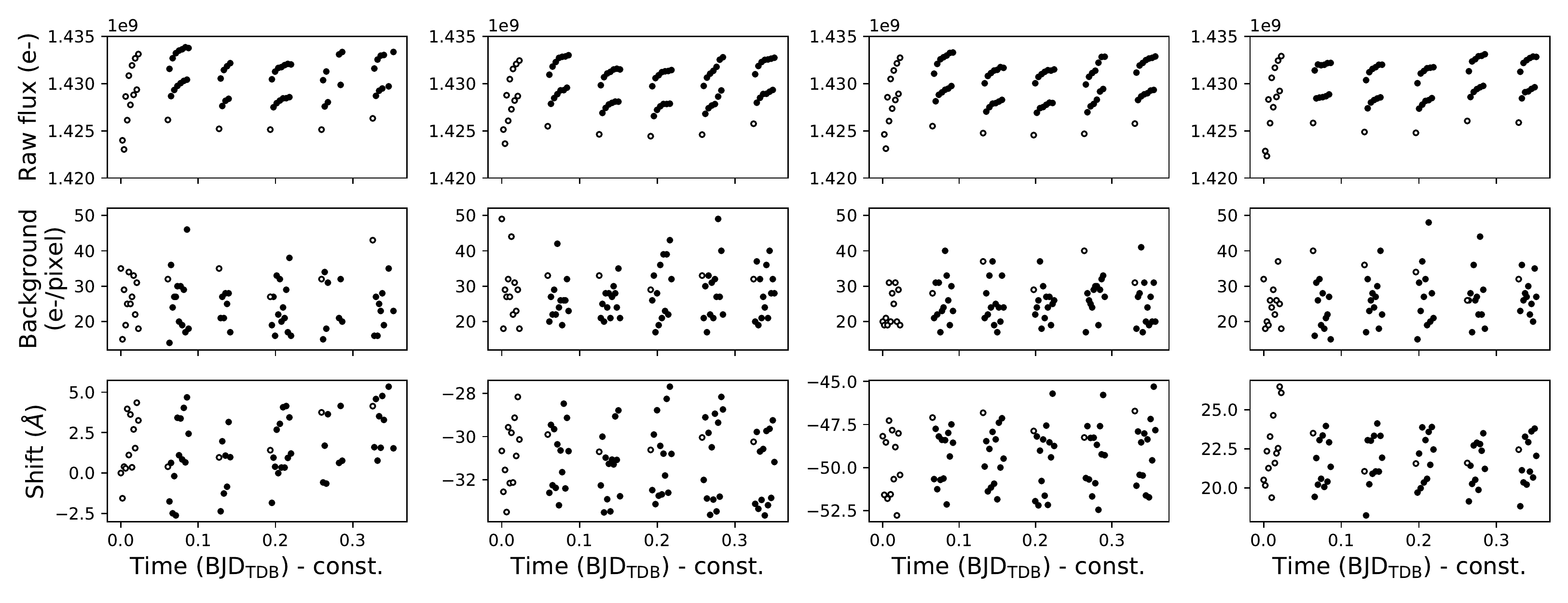}
\caption{Diagnostic information from the \HST\ data reduction. From top to bottom, the rows show the band-integrated raw flux, the background counts, and the wavelength shift of the spectrum relative to the first exposure of the first visit. From left to right, the columns show the four \HST\ visits in chronological order. The open circles in the raw flux correspond to data points that we did not include in our light curve fits due to  larger amplitude instrumental ramps. The vertical offset in the top row is due to spatial scanning, which alternates between forward and reverse directions on the detector. The total counts are higher when the detector is read out in the same direction as the spatial scan.}
\label{fig:diagnostics}
\end{center}
\end{figure*}

We used a custom data reduction pipeline to process the \HST\ transit observations \citep[described in detail in][]{kreidberg18b}. The starting point for our reduction was the \texttt{\_ima} data product provided by the Space Telescope Science Institute. These images have an intermediate level of processing, with corrections applied for dark current, linearity, and flat fielding. To extract spectra from the images, we used the optimal extraction routine of \citep{horne86}. This algorithm iteratively masks bad pixels in the image, and provides a convenient method to reject cosmic rays from spatial scan data. To estimate the background, we identified a region of pixels that was not contaminated by flux from the target or any nearby stars (rows $10-70$ and columns $400 - 500$) and calculated the median count rate in this region. We subtracted the background and extracted the spectra from each up-the-ramp sample separately, and summed them to produce a final spectrum from the exposure. To account for spectral drift, we interpolated each spectrum to the wavelength scale of the first exposure of the first visit. We generated spectroscopic light curves by binning the spectra into 22 wavelength channels over the wavelength range $1.125 -1.65\,\mu$m. This binning corresponds to roughly five pixels in the spectral direction. The binning is about twice as coarse as the native resolution of the grism, and was chosen to average over variations in sensitivity between individual pixels.   Figure\,\ref{fig:diagnostics} shows the band integrated light curve, the background counts, and the spectral shifts for each visit.

We fit the light curves with a joint model of the transit and the instrument systematic trends. In agreement with previous work, we found that the first orbit of every visit and the first exposure in each orbit were strongly affected by a ramp-like systematic (caused by charge traps filling up in the detector; \citealt{zhou17}). This systematic is visible in the raw data, shown in Figure\,\ref{fig:diagnostics}. Following past studies, we removed the first orbit of the visit and the first exposure of the remaining orbits in our analysis \citep[e.g.][]{kreidberg14a}. The trimmed data had three full orbits per visit of out-of-transit baseline, which is sufficient to fit for visit-long trends.

\begin{figure*}
\begin{center}
\includegraphics[scale = 1.0]{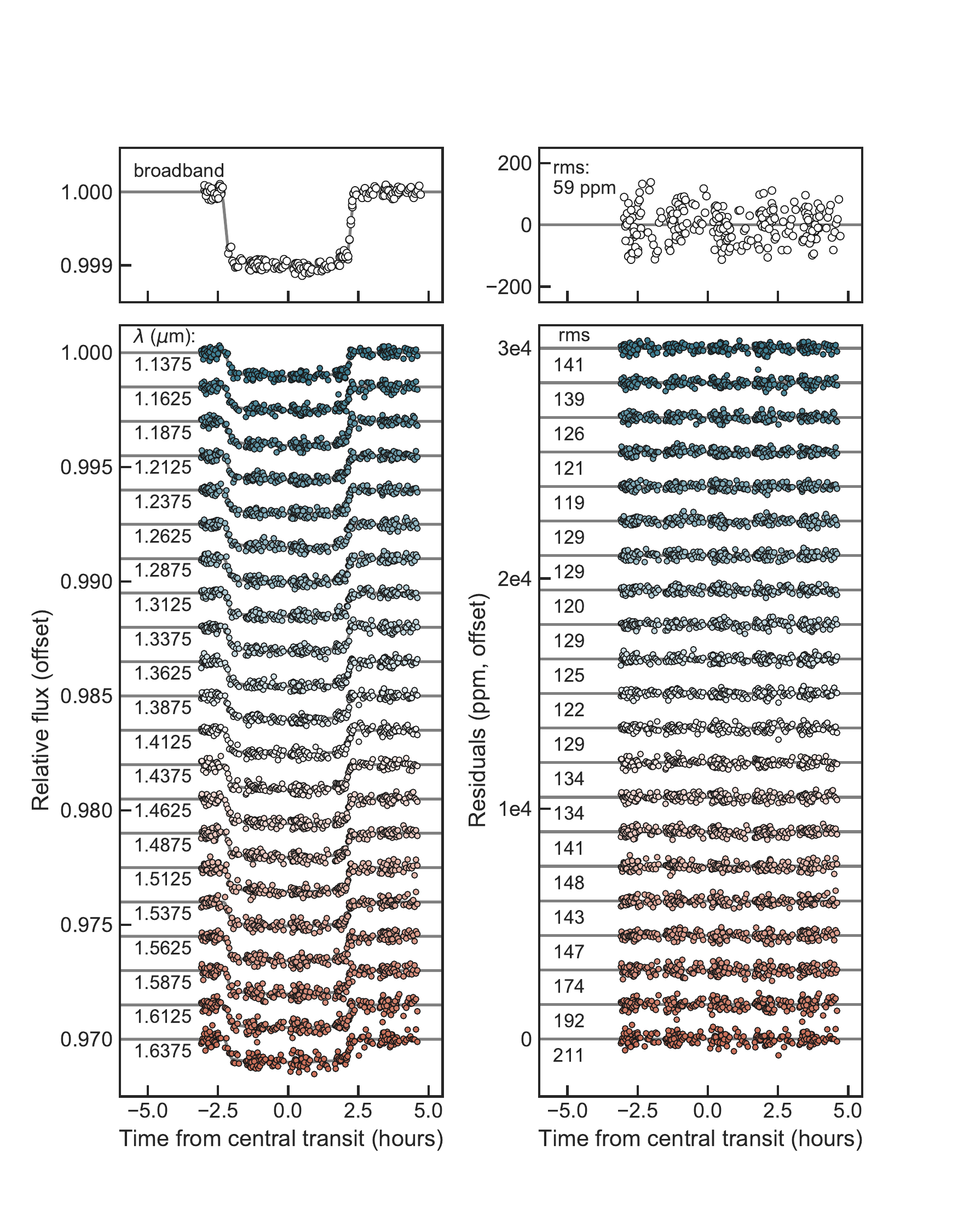}
\caption{HST/WFC3 transit light curves. The left panel shows the phase-folded light curve from all four HST visits (points) compared to the best fit models (lines) for the broadband light curve (top) and the spectroscopic light curve (bottom). The right panel shows the residuals from the best fit models (right). The figure is annotated with the central wavelength for each spectroscopic light curve and the rms of the residuals (in ppm). The data are corrected for instrument systematics, normalized to a out-of-transit baseline flux of unity, and offset on the y-axis by a constant value for visual clarity.}
\label{fig:hst_lc}
\end{center}
\end{figure*}

\begin{figure*}
\begin{center}
\includegraphics[scale = 0.7]{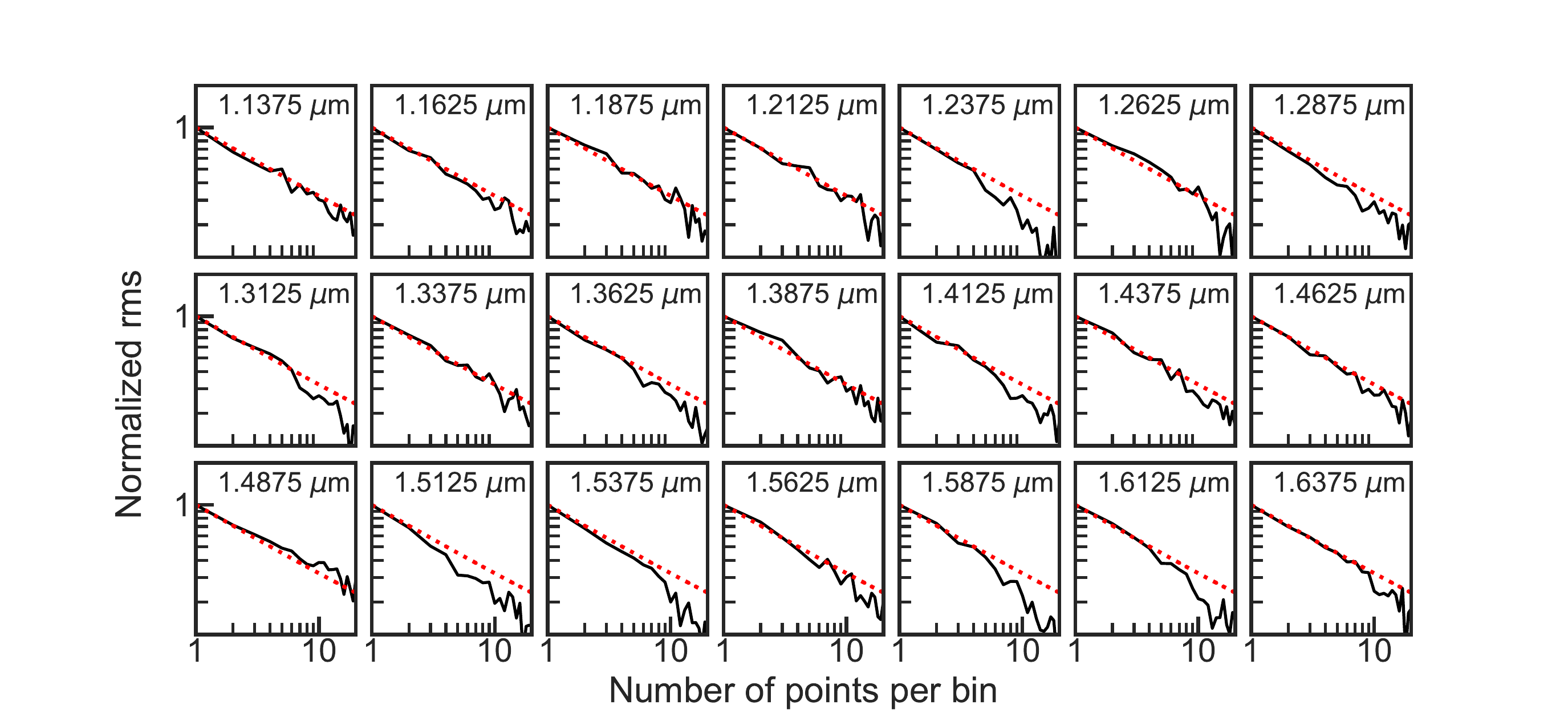}
\caption{Allan deviation plots showing the rms variation as a function of bin size for the HST/WFC3 transit light curves. A bin size of ten points corresponds to 30 minutes. The red dotted lines show the expected $\sqrt{N}$ trend for photon-limited, white noise. The black lines show the measured rms for each light curve.}
\label{fig:allan}
\end{center}
\end{figure*}

To model the transit signal, we used the \texttt{batman} package \citep{kreidberg15a}. For the broadband light curve fit, the free parameters for the transit model were the ratio of planet to stellar radius $R_p/R_s$, the time of central transit $T_c$, the orbital inclination $i$, the ratio of semi-major axis to stellar radius $a/R_s$. We fixed the eccentricity to zero. We ran an initial fit with free parameters for a linear limb darkening coefficient, and found excellent agreement with predictions from a Kurucz ATLAS9 stellar model with $T_\mathrm{eff}=6250$K, $\mathrm{log}\ g = 4.5$ (cgs), and [Fe/H]=$-0.2$ \citep{castelli03}. We therefore fixed the limb darkening on the predicted quadratic coefficients from the model for the remainder of the analysis\footnote{calculated with the ExoCTK Limb Darkening Calculator; \url{https://exoctk.stsci.edu/}}. For the spectroscopic channels, we fixed $T_c$, $a/R_s$, and $i$ on the best-fit values from the broadband light curve.

To model systematic noise from the instrument, we multiplied the model transit light curve by the analytic \texttt{model-ramp} function, previously used for WFC3 data analysis \citep[Equation 3;][]{kreidberg18b}. Briefly, this function fits an exponential ramp to each orbit, and a visit-long trend. For the \planet\ data, there was no significant improvement to the light curve fit for a quadratic term in the visit-long trend, so we used a linear term only.  The light curves and best-fit models are shown in Figure~\ref{fig:hst_lc}. To test for correlated noise in the light curve, we binned the data in time over a range of bin sizes from $2 - 20$ points and calculated the rms for each bin size (see Figure~\ref{fig:allan}). The rms decreases with the square root of the number of points per bin, indicating that the noise is uncorrelated in time.

We used the \texttt{dynesty} package to estimate parameter uncertainties for our model \citep{speagle20}. The package uses dynamic nested sampling to evaluate constant likelihood contours over the full prior volume. To ensure that we did not underestimate the uncertainties, we rescaled the per point errors on the data such that the reduced $\chi^2$ of the best fit model was unity. The error bars increased by a median (mean) of 9\% (12\%). The \texttt{dynesty} runs were halted when the remaining contribution to the evidence was estimated to be below 0.01 of the total.  The resulting median and 68\% credible intervals for the transit depths are listed in Table\,\ref{tab:spec_lc_params}.

\subsubsection{Independent Analysis of the WFC3 Data}
We also carried out an independent data reduction and analysis. The data were reduced following the methodology previously described by \cite{evans16, evans17}. Briefly, the spectra are extracted from each \texttt{\_ima} frame by summing the difference of successive up-the-ramp samples while masking cross-dispersion regions away from the target to reject cosmic rays and nearby contaminating sources. A wavelength-independent background value was subtracted from each spectrum by taking the median pixel value in a $30 \times 250$ pixel box away from the target. Broadband light curves were produced for each visit by summing each spectrum along the full dispersion axis. The broadband light curves were fit jointly, with the systematics and transit mid-times allowed to vary separately, and $R_p/R_s$ shared across visits. Other transit parameters such as $a/R_s$ and $i$ were fixed to the median values reported in \cite{crossfield17b}. For the systematics, we adopted the double-exponential ramp treatment described in \cite{dewit18} and also allowed the white noise to vary as a free parameter, implemented as an increase above the formal photon noise value. 

Following the broadband light curve fit, we produced spectroscopic light curves in 14 channels spanning the $1.122-1.642\mu$m wavelength range, following the methodology described in \cite{evans16}, which is based on an original implementation of \cite{deming13}. This procedure effectively removes the common-mode component of the systematics in each wavelength channel, which is dominated by the ramp systematic. As such, for our spectroscopic light curve fits, a simple linear time trend and variable white noise level were adequate for modeling the systematics. We also allowed $R_p/R_s$ to vary, while holding all other transit parameters fixed to the white light curve values. In both the white light curve and spectroscopic light curve fits, a quadratic limb darkening law was adopted with coefficients held fixed to values determined by fitting the limb darkened profiles of an ATLAS9 stellar model with the same parameters listed in~\ref{sec:wfc3}. 

\begin{figure}
\begin{center}
\includegraphics[scale = 0.6]{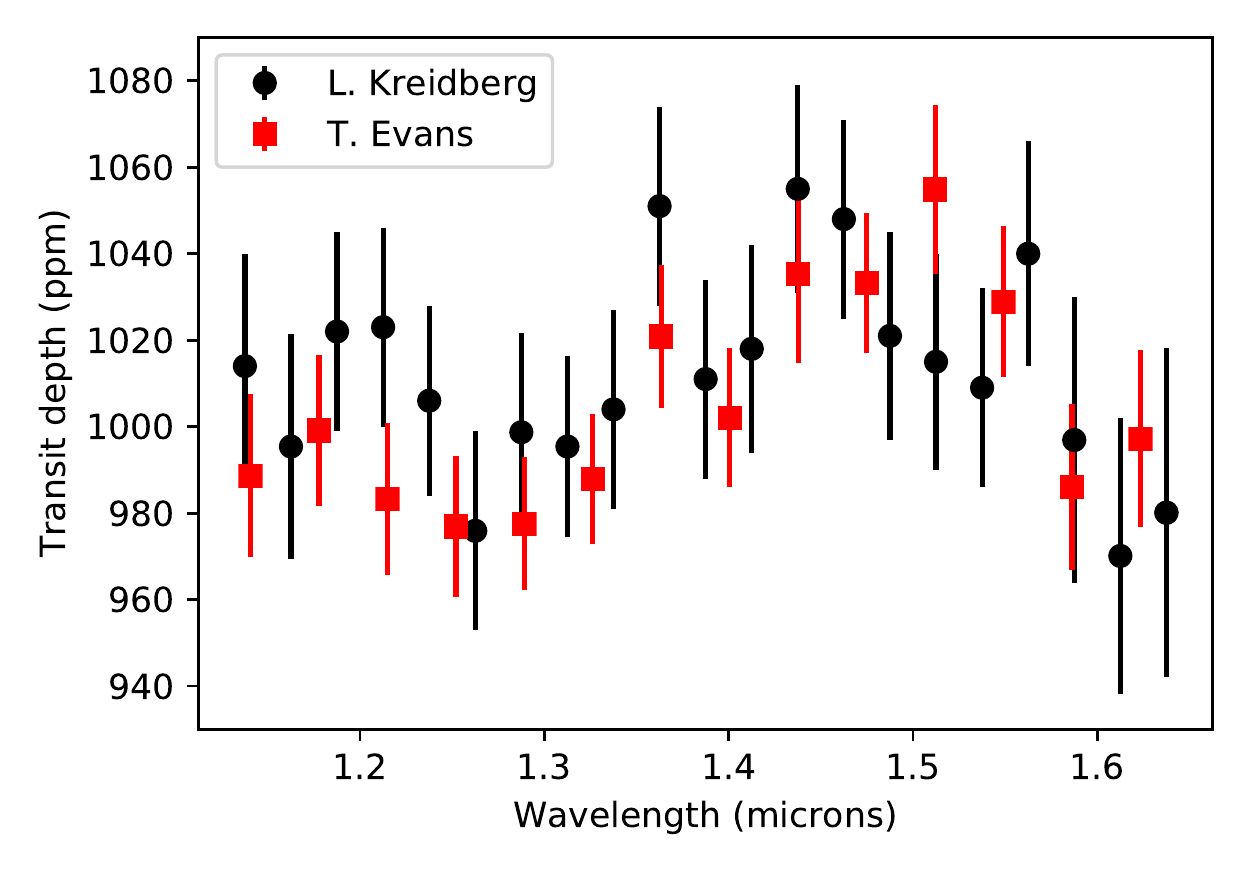}
\caption{\HST/WFC3 transmission spectra from two independent pipelines. The black circles are from the \texttt{model-ramp} analysis used by L. Kreidberg, whereas the red squares come from the common-mode error correction from T. Mikal-Evans.}
\label{fig:compare}
\end{center}
\end{figure}

The resulting transmission spectrum is compared to the first analysis  in Figure~\ref{fig:compare}. The two spectra agree to well within $1\sigma$, and the uncertainties on the transit depths are consistent after accounting for the difference in wavelength bin size. The \texttt{model-ramp} fit has a median uncertainty on the transit depth of 23 ppm (for $0.025\,\mu$m bins), and the common-mode fit has a median uncertainty of 17 ppm (for $0.037\, \mu$m bins). Given the good agreement between the two independent analyses, we use the \texttt{model-ramp} results (listed in Table\,\ref{tab:spec_lc_params}) for the remainder of the analysis.

\begin{table}[]
\begin{tabular}{llll}
Wavelength & ($R_p/R_s)$$^2$ & $u_1$ & $u_2$\\
($\mu$m) & (ppm) & (fixed) & (fixed) \\
\hline
\hline
$ 0.42 - 0.9 $ & $ 1030 \pm 26 $ & 0.365 & 0.244 \\
$ 1.125 - 1.150 $ & $ 1014 \pm 26 $ & 0.180 & 0.214 \\
$ 1.150 - 1.175 $ & $ 995 \pm 26 $ & 0.177 & 0.214 \\
$ 1.175 - 1.200 $ & $ 1022 \pm 23 $ & 0.171 & 0.214 \\
$ 1.200 - 1.225 $ & $ 1023 \pm 23 $ & 0.169 & 0.215 \\
$ 1.225 - 1.250 $ & $ 1006 \pm 22 $ & 0.166 & 0.215 \\
$ 1.250 - 1.275 $ & $ 976 \pm 23 $ & 0.162 & 0.215 \\
$ 1.275 - 1.300 $ & $ 999 \pm 23 $ & 0.155 & 0.217 \\
$ 1.300 - 1.325 $ & $ 995 \pm 21 $ & 0.132 & 0.230 \\
$ 1.325 - 1.350 $ & $ 1004 \pm 23 $ & 0.148 & 0.218 \\
$ 1.350 - 1.375 $ & $ 1051 \pm 23 $ & 0.145 & 0.216 \\
$ 1.375 - 1.400 $ & $ 1011 \pm 23 $ & 0.140 & 0.217 \\
$ 1.400 - 1.425 $ & $ 1018 \pm 24 $ & 0.136 & 0.216 \\
$ 1.425 - 1.450 $ & $ 1055 \pm 24 $ & 0.132 & 0.215 \\
$ 1.450 - 1.475 $ & $ 1048 \pm 23 $ & 0.129 & 0.213 \\
$ 1.475 - 1.500 $ & $ 1021 \pm 24 $ & 0.123 & 0.214 \\
$ 1.500 - 1.525 $ & $ 1015 \pm 25 $ & 0.116 & 0.214 \\
$ 1.525 - 1.550 $ & $ 1009 \pm 23 $ & 0.112 & 0.212 \\
$ 1.550 - 1.575 $ & $ 1040 \pm 27 $ & 0.108 & 0.208 \\
$ 1.575 - 1.600 $ & $ 997 \pm 33 $ & 0.102 & 0.205 \\
$ 1.600 - 1.625 $ & $ 970 \pm 32 $ & 0.096 & 0.204 \\
$ 1.625 - 1.650 $ & $ 980 \pm 38 $ & 0.091 & 0.201 \\
$ 4.0 - 5.0 $ & $ 1070 \pm 72 $ & 0.079 & 0.089 \\
\end{tabular}
\caption{Transit depths and limb darkening coefficients for the \Ktwo, \HST, and \Spitzer\ data. The transit depth values are the median and 68\% credible interval from the posterior distributions. The limb darkening parameters are pre-calculated from stellar models and fixed in the analysis.}
\label{tab:spec_lc_params}
\end{table}

\subsection{Spitzer}
In addition to the HST and K2 data, we also analyzed a single transit of \planet\ observed with {\em Spitzer} in the 4.5 $\mu$m bandpass.  We follow a similar approach as the one described in \cite{berardo19}, which detrends the data using the Pixel Level Decorrelation method outlined in  \cite{deming15}. We first applied a median filter to each pixel in the image and calculated a background level for each frame by taking the median of the flux in an annulus centered on the point spread function. We estimated the centroid of each frame by fitting a two dimensional Gaussian to the image, and obtained a light curve using a fixed radius aperture. 

\begin{figure*}
\begin{center}
\includegraphics[scale = 0.7]{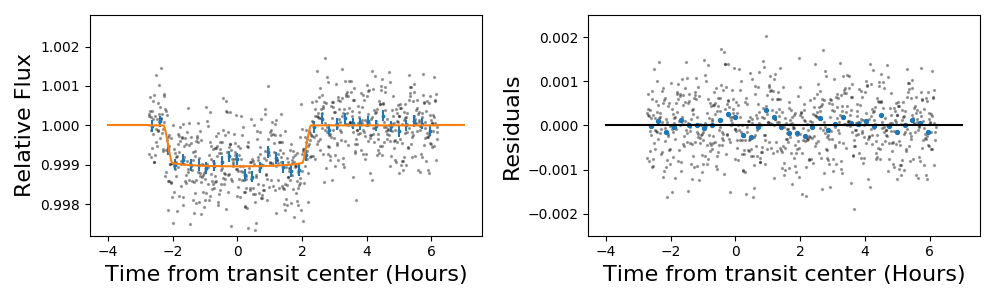}
\caption{The light curve of \planet\ observed with the 4.5 $\mu$m filter of {\em Spitzer}. The left panel shows the best fit transit model to the binned light curve after removing detector systematics. The blue points with error bars are the data points binned further for visual clarity. The right hand panel shows the residuals of the best fit model from the data. }
\label{fig:spitzer transit}
\end{center}
\end{figure*}

 We modeled systematics in the light curve by weighting the nine brightest pixels individually as well as fitting for a quadratic time ramp. We then chose the combination of pixel coefficients, aperture size, and time-series binning that resulted in the smallest root mean square (rms) deviation. The optimal aperture radius was found to be 2.4 pixels. We used an MCMC sampler to estimate uncertainties, and fit the systematic model simultaneously with a transit  model from \texttt{batman} \citep{kreidberg15a}. We kept the period, inclination, and distance $a/R_\star$ fixed to the values 21.0564 days, $88.501^{\circ}$, and 26.769, respectively (based on the \HST\ white light curve fit), and allowed the depth and transit center to vary. We also left the uncertainty of the data points as a free parameter, which we found converged to the rms scatter of the raw light curve. We also held fixed the quadratic limb darkening parameters, which were also estimated from a Kurucz ATLAS9 stellar model. The transit light curve and best fit model are shown in Figure~\ref{fig:spitzer transit}, and the fit results are summarized in Table~\ref{tab:ktwotransit}. 

\subsection{K2}

 To provide a broadband, optical-wavelength transit depth for comparison with our infrared observations, we reanalyzed the 30-minute-cadence \Ktwo\ photometry of HD~106315. Although several analyses of these \Ktwo\ data have already been published \citep{crossfield17b,rodriguez17}, our reanalysis takes advantage of the tighter constraints on orbital parameters ($a/R_s$ and $i$) provided by the higher-cadence {\em Spitzer} and {\em HST} observations.
Our analysis used largely the same approach described by \citep{crossfield17b}, but with a few changes. First, we used a different set of \Ktwo\ photometry\footnote{\url{https://www.cfa.harvard.edu/~avanderb/k2.html}} which had a substantially lower rms.  Second, we fixed two key orbital parameters to the following values: $a/R_s=26.769$, and $i=88.501^{\circ}$. Third, in contrast with the analysis of \cite{crossfield17b} we allowed no dilution that would potentially decrease the observed transit depth (and so bias the analysis toward larger $R_p/R_s$). We neglected dilution because high-resolution imaging and spectroscopy show no nearby stars within 5 magnitudes of HD~106315 at distances $<0.1$ arcsec (Kosiarek et al., in prep). Finally, we held the quadratic limb darkening parameters fixed to the values predicted by an ATLAS9 stellar model ($u_1=0.365$ and $u_2=0.244$).  The transit parameters derived from this analysis are listed in Table~\ref{tab:ktwotransit}.

\begin{deluxetable*}{l l l l}[bt]
\tabletypesize{\scriptsize}
\tablecaption{ \Ktwo / {\em Spitzer} Transit Parameters \label{tab:ktwotransit}}
\tablewidth{0pt}
\tablehead{
\colhead{Parameter} & \colhead{Units} & \colhead{Value (K2)}  & \colhead{Value (Spitzer)}
}
\startdata
       {\em Held fixed:} & & & \\
   $R_s/a$ &         -- & $0.0373566$ & $0.0373566$ \\
       $i$ &        deg & $88.50109$ & $88.50109$\\ 
       $u_1$ &         -- & $0.365$ & $0.079$\\
       $u_2$ &         -- & $0.244$ & $0.089$ \\
   {\em Derived values:} & & \\
   $T_{0}$ & $BJD_{TDB} - 2454833 $ & $2778.1320^{+0.0016}_{-0.0017}$ & $3030.8079\pm 0.0012$ \\
       $P$ &          d & $21.0564 \pm 0.0024$ & $21.0564\ $(fixed) \\
 $R_p/R_s$ &         \% & $3.208 \pm 0.041$ & $3.271 \pm 0.11$ \\
 $(R_p/R_s)^2$ &         ppm & $1030 \pm 26$ & $1070 \pm 75$\\
\enddata
\end{deluxetable*}

\subsection{Potential Impact of Star Spots}
Unocculted star spots and plages can significantly contaminate exoplanet transmission spectra \citep{pont13, rackham18, rackham19}. In general, F-stars like HD~106315  have lower spot covering fractions than stars of later spectral type \citep{rackham19}. The \Ktwo\ light curve for HD~106315 shows variability with amplitude 0.1\% over a timescale of 75 days, a typical value for mid-F stars \citep{rodriguez17}. This amplitude corresponds to a covering fraction of $0.1\pm0.1\%$. The expected amplitude of the stellar contamination spectrum is $0.0001 - 0.0002\times$ the transit depth. For the transit depth of \planet\ (1000 ppm), the expected stellar contamination is $0.1 - 0.2$ parts per million --- a negligible contribution. 

\section{Atmospheric Retrieval} \label{sec:retrieval}
We carried out two independent retrieval analyses to determine the molecular abundances and cloud properties of the \planet's atmosphere. We used the open-source software package \texttt{petitRADTRANS} (pRT) \citep{molliere19}, as well as a retrieval based on the \texttt{SCARLET} framework and \citep{benneke13,benneke19}. Both retrieval analyses used a Bayesian framework to compare the measured spectrum to one-dimensional models with variable atmospheric properties as described in this section. The analyses consistently provide tentative evidence for water vapor based on Bayesian model comparison of retrieval models \citep{benneke13}.

\subsection{\texttt{petitRADTRANS} retrieval analysis}
 \label{sec:PTRTretrieval}
We used the open-source software package \texttt{petitRADTRANS} (pRT) \citep{molliere19}, which is a fast spectral synthesis tool for exoplanet atmospheres.  We connected pRT to the \texttt{PyMultiNest} tool \citep{buchnergeorgakakis2014}, which is a Python wrapper of the MultiNest \citep{ferrozhobson2008,ferrozhobson2009,ferrozhobson2013} implementation of nested sampling \citep{skilling2004}.

The atmosphere was modeled with the vertically constant temperature and absorber mass fractions of H$_2$O, CH$_4$, CO$_2$, CO and N$_2$ as free parameters. We also included the cloud top pressure of a gray cloud deck as a free parameter. The atmospheric mean molecular weight (MMW) was calculated from the parameterized absorber abundances, assuming that the remaining mass is contributed by H$_2$ and He, with a H$_2$:He mass ratio of 3:1. Our full model included the line opacities of H$_2$O, CH$_4$, CO$_2$ and CO, as well as the Rayleigh scattering cross-sections of these species, in addition to those of H$_2$, He and N$_2$.
N$_2$ may thus be thought of as a proxy for (mostly) invisible species in the atmosphere that can increase its MMW. Instead of retrieving a reference radius at a given pressure we retrieved a reference pressure $P_0$ at a given radius, where we made sure that the fixed reference radius is chosen at values appropriate for placing the retrieved reference pressure values within the atmospheric pressure domain. We placed log-uniform priors on the absorber mass fractions of H$_2$O, CH$_4$, CO$_2$, CO and N$_2$ between $10^{-10}$ and 1, requiring that the sum of all mass fractions is below unity. The temperature was allowed to vary between 400 and 1000~K. The cloud and reference pressure could be placed at any location within the atmospheric pressure domain, imposing a log-uniform prior. However, we note that the posterior distribution for the water abundance (which is the only species we detect tentatively) is sensitive to the choice of prior bounds, particularly if regions are explored that do not produce any difference in the model spectrum (for example, very deep clouds).   We additionally tested retrieving a scattering haze ($\kappa_{\rm haze} = \kappa_0 [\lambda/\lambda_0]^\gamma$), a cloud patchiness parameter (mixing clear and cloudy terminators), and the planet's gravity within measurement uncertainties, but none of these tests significantly changed our results.

\begin{figure*}
\begin{center}
\includegraphics[scale = 0.6]{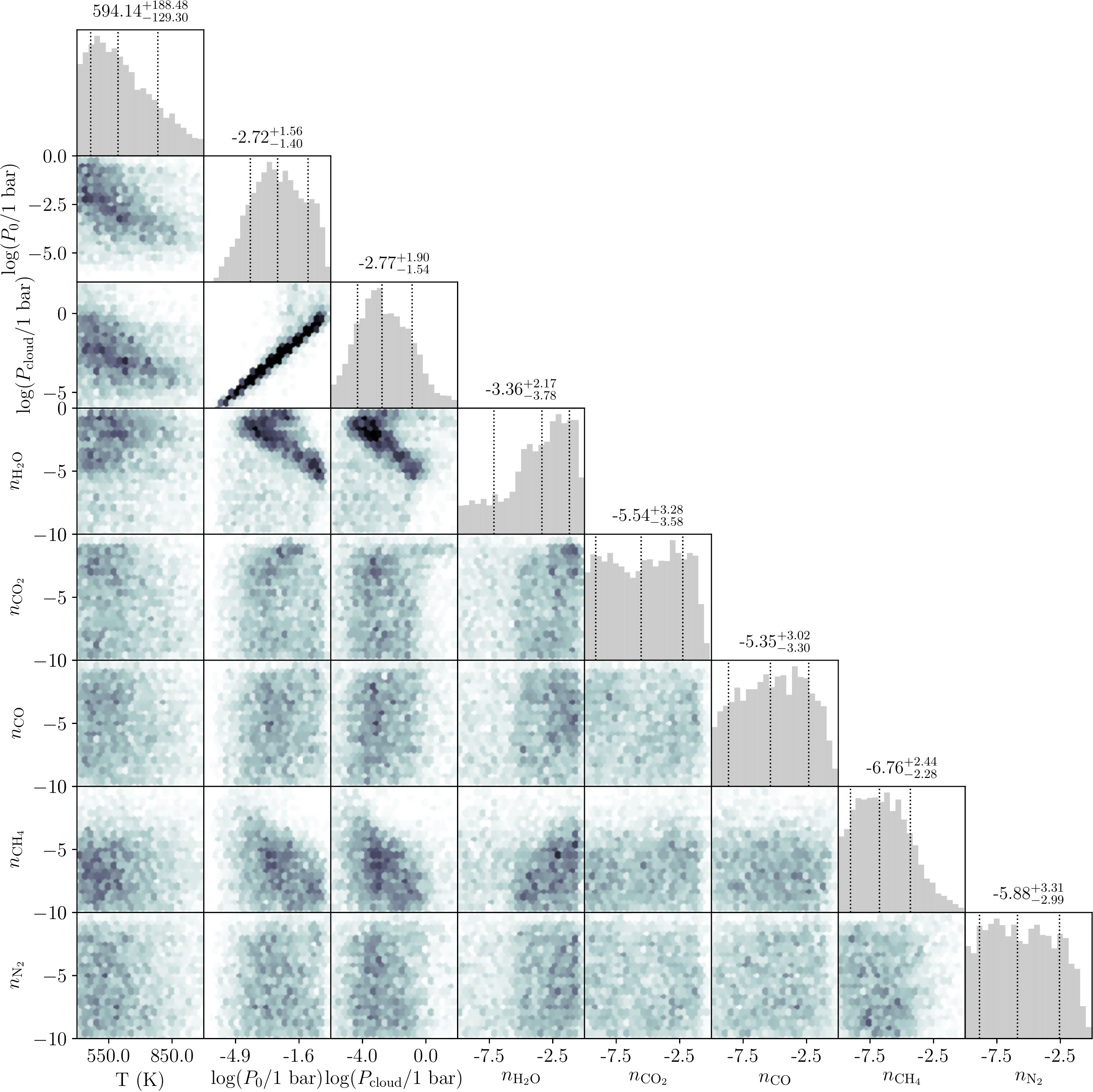}
\caption{The retrieved atmospheric properties for the full retrieval with petitRADTRANS. The panels show the posterior distribution of parameters from the nested sampling run. Darker shading corresponds to higher posterior probability. The diagonal shows a 1-dimensional histogram for each parameter, with dotted lines denoting the median and $1\sigma$ credible interval. The molecular abundances are the logarithm (base 10) of their volume mixing ratio. For reference, a solar composition gas at 1 mbar pressure and 800 K has $n_\mathrm{H_2O} = -3.65$, $n_\mathrm{CO2} = -6.32$, $n_\mathrm{CO} = -3.37$, $n_\mathrm{CH_4} = -4.43$ and $n_\mathrm{N_2} = -4.23$.}
\label{fig:corner}
\end{center}
\end{figure*}


\begin{table}[]
\caption{Bayesian Evidence for Atmospheric Retrievals with petitRADTRANS}
\begin{tabular}{lll}
Retrieval model      & $\Delta\,ln(Z)$      & Bayes factor for     \\
                     &                      & molecule present     \\
\hline
\hline
no H$_2$O & $  -0.517 \pm 0.078 $ & \bf{1.68} \\
no CO$_2$ & $  0.347 \pm 0.085 $ & 0.71 \\
no CO & $  0.490 \pm 0.122 $ & 0.61 \\
no CH$_4$ & $  0.482 \pm 0.040 $ & 0.62 \\
no N$_2$ & $  0.766 \pm 0.237 $ & 0.46 \\
no cloud & $  -0.022 \pm 0.067 $ & 1.02 \\
\end{tabular}
\label{tab:retrieval}
\end{table}

\begin{table}[]
\caption{Bayesian Evidence for SCARLET Atmospheric Retrievals}
\begin{tabular}{lll}
\hline
\hline
Retrieval model      & $\Delta\,ln(Z)$      & Bayes factor for     \\
                     &                      & molecule present     \\
\hline
no H2O               & $+0.961\pm0.010$     & \bf{2.61} (1.9$\sigma$)   \\
no CO2               & $+0.010\pm0.028$     & 1.01                 \\
no CO                & $+0.022\pm0.028$     & 1.02                 \\
no CH4               & $-0.382\pm0.026$     & 0.68                 \\
no N2                & $-0.020\pm0.028$     & 0.98                 \\
no clouds            & $+0.661\pm0.038$     & 1.94                 \\
\hline
\end{tabular}
\label{tab:scarlet_bayesfactor}
\end{table}

In order to test how reliably water can be detected in our spectrum we followed the approach introduced in \citet{benneke13}. Our full model retrieved the abundance of all absorbers listed above. Then we iteratively removed one absorber at a time and re-ran the retrieval. Comparing the evidences between the full model and the model lacking a given species allows to assess whether the observation is in favor of that species being included in the model. The retrieved atmospheric properties for the full model are shown in Figure~\ref{fig:corner}.
 In addition to the "full retrieval model", which includes all five molecules (H$_2$O, CH$_4$, CO$_2$, CO and N$_2$), we ran five additional retrieval models, each with one molecular species removed at a time. This approach of removing one molecular species from full model at a time enables us to unambiguously check for each individually molecule and captures any ambiguity that may be introduced by overlapping absorption features \citep{benneke13}.

The resulting evidences $Z$ are listed in Table~\ref{tab:retrieval}. The Bayes factor is the ratio of evidences.  Bayes factors greater than 100 are considered decisive, $10 - 100$ is strong, $3.2 - 10$ is substantial, and below 3.2 is insignificant \citep{kass95}.
None of the tested species is substantially favored to be included in our model. However, the full model is slightly favored when compared to a model that removed H$_2$O, with a Bayes factor of 1.7.  The fact that water is the only molecule that can lead to noticeable differences in the fit is not surprising, since WFC3 spectra are predominantly sensitive to absorption from H$_2$O.  Higher precision measurements of the transmission spectrum are needed to uniquely identify absorbing species in the atmosphere of HD~106315~c.

\begin{figure*}
\begin{center}
\includegraphics[scale = 0.7]{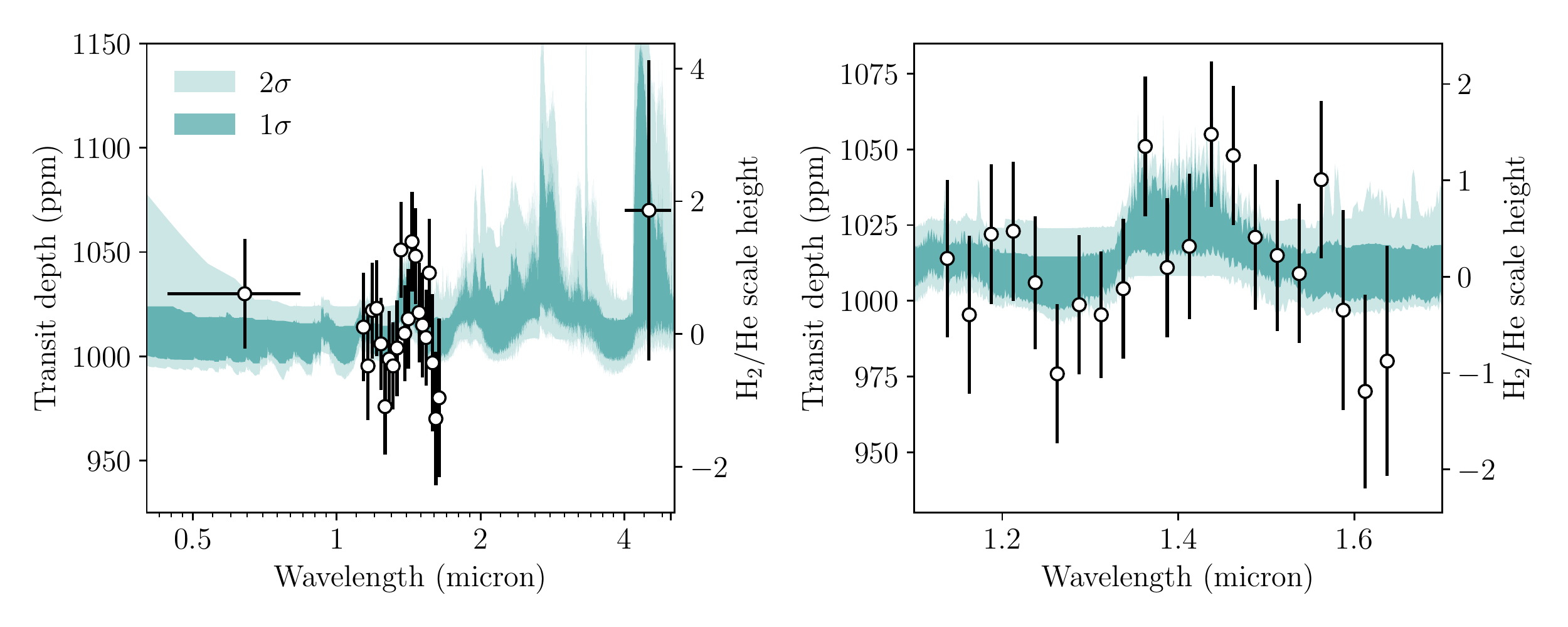}
\caption{The transmission spectrum of \planet\ (points with 1$\sigma$ uncertainties) compared to retrieved spectra from the FULL model (teal shading) with petitRADTRANS. The H$_2$/He atmospheric scale height is indicated on the right y-axis, assuming a solar composition atmosphere at the planet's equilibrium temperature (the true scale height is likely smaller, due to enhanced metallicity and/or lower temperature). The tentative detection of water absorption is driven by the small increase in transit depth near $1.4\mu$m.}
\label{fig:retrieved_spectra}
\end{center}
\end{figure*}

\begin{figure*}
\begin{center}
\includegraphics[width = 0.9\textwidth]{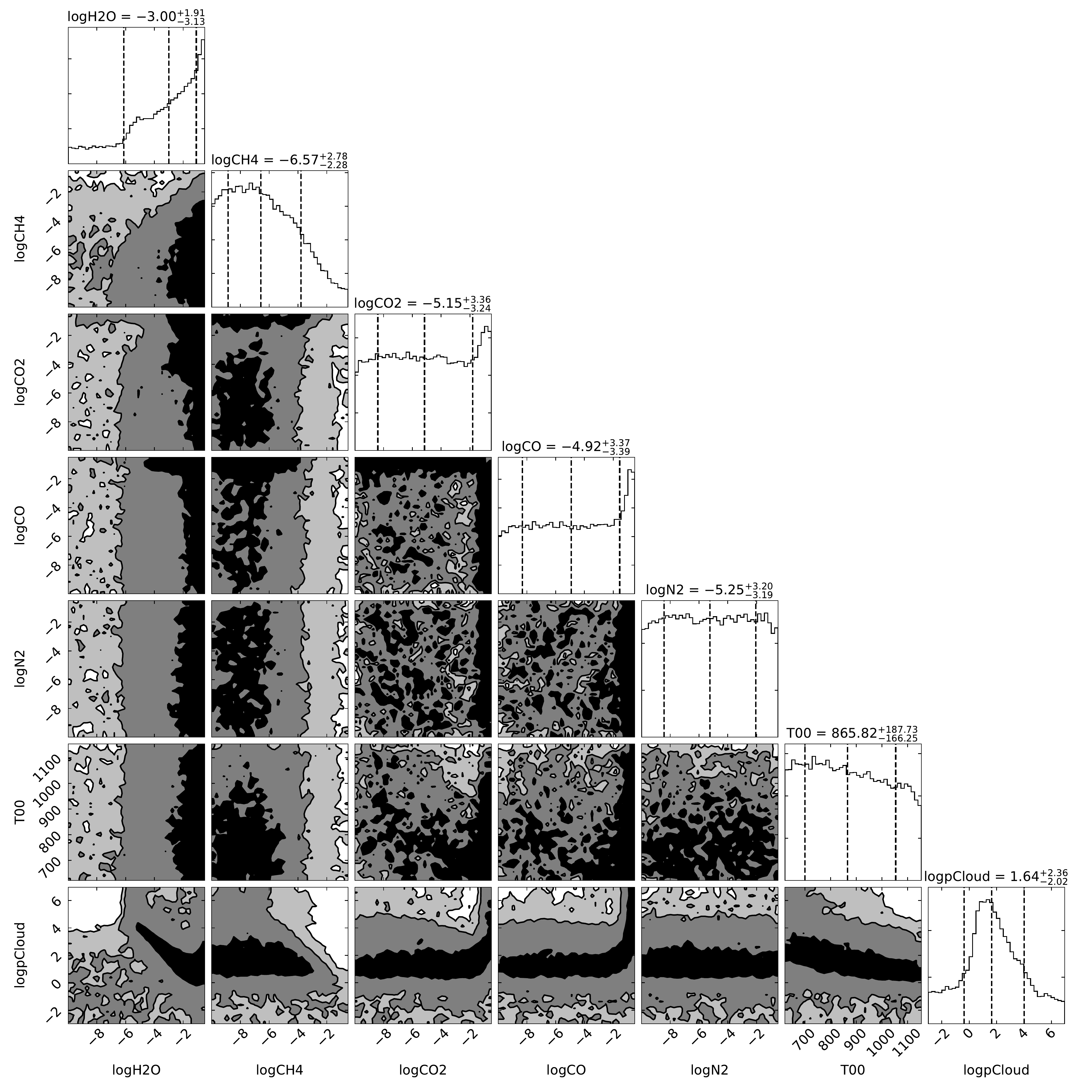}
\caption{Molecular abundance and cloud property constraints from the SCARLET free retrieval analysis. The top panels in each column show the 1D marginalized posterior distributions of the molecular abundances and cloud properties, with dashed vertical lines in the histograms indicating the marginalized 16th, 50th, and 84th credible intervals. The subjacent 2D panels show the correlations among the gases and cloud properties, with the black, dark-gray, and light-gray regions in corresponding to the 1$\sigma$ (39.3\%), 2$\sigma$ (86.5\%), and 3$\sigma$ (98.9\%) credible intervals. The cloud pressure is given in logarithm (base 10) Pascals, and the abundances are logarithm (base 10) of their volume mixing ratio.}
\label{fig:scarlet_cornerplot}
\end{center}
\end{figure*}

The retrieved spectrum is shown in Figure~\ref{fig:retrieved_spectra}.  The amplitude of spectral features in the best fit model is about 30 ppm ($7\times$ smaller than that expected for a  solar composition, cloud-free atmosphere). This observed peak-to-trough amplitude  corresponds to $0.8 \pm 0.04$ H$_2$/He scale heights (assuming $\mu = 2.3$ atomic mass units, $T = 800$~K, and $g = 6.0$~m/s$^2$). In general, to produce features of this amplitude, models have either (1) enhanced mean molecular weight (which decreases the atmospheric scale height and shrinks the spectral features), (2) high altitude clouds, which truncates the spectral feature at the cloud deck altitude, or (3) both of the above \citep{benneke13}. In the case of \planet, the retrieval prefers scenarios (2) and (3), with the highest posterior probability for moderate cloud coverage. A broad range of H$_2$O abundances are consistent with the data ($3\times10^{-4} - 290\times$ solar at 1$\sigma$). The ``solar" water abundance corresponds to the chemical equilibrium water volume mixing ratio for a solar composition gas at 1 mbar pressure and 800 K ($2.2\times10^{-4}$). The cloud-top pressure is between $P_\mathrm{cloud} = 0.04 - 130$ mbar (at 1 $\sigma$). There is some degeneracy between $n_\mathrm{H_2O}$ and $P_\mathrm{cloud}$, because higher water abundance pushes the photosphere to lower pressures. There is also a tail of probability towards water-rich solutions with deep clouds (below the observable photosphere).   Very high water abundances cannot be ruled out ($n_\mathrm{H_2O}<2100\times$ solar at 2$\sigma$ confidence, $n_\mathrm{H_2O} < 4200\times$ solar at 3$\sigma$).
 
\subsection{SCARLET retrieval analysis}
As an independent check of the results from petitRADTRANS, we also interpreted the transmission spectrum with  the SCARLET atmospheric retrieval framework \citep[e.g.,][]{benneke12,benneke13,kreidberg14a,knutson14a,benneke15,benneke19a,benneke19,wong_optical_2020}. Employing SCARLET's free molecular composition mode we defined the mole fractions of H$_2$O, CH$_4$, CO$_2$, CO and N$_2$ as free parameters and assumed a well-mixed atmosphere. The remainder of the atmosphere gas was assumed to be composed of H$_2$ and He in solar abundance ratio. We included a gray cloud deck using a free parameter describing the cloud top pressure, and an additional free parameter to capture our prior ignorance of  the temperature in the photosphere of \planet\  near the terminator. 

To evaluate the likelihood for a particular set of atmospheric parameters, the SCARLET forward model in free molecular composition mode computes the hydrostatic equilibrium and line-by-line radiative transfer. We consider the latest line opacities of H$_2$O, CO, and CO$_2$ from HiTemp \citep{rothman} and CH$_4$ from ExoMol \citep{tennyson16}, as well as the collision-induced absorption of H$_2$ and He. We employed log-uniform priors between $10^{-10}$ and $10^{-0.5}=31\%$, but required that the sum of all mass fractions is below unity. We employed log-uniform priors for the cloud top pressures  $10^{-3}$ and $10^7$ Pa. We used a uniform prior on the photospheric temperature between 620K and 1150K (70--130\% of the equilibrium temperature).

SCARLET then determined the posterior constraints by combining the SCARLET atmospheric forward model with nested sampling \citep{skilling2004}. We ran the analyses well beyond formal convergence to obtain smooth posterior distribution and capture the contours of the wide parameter space in agreement with the transmission spectrum of \planet. As in Section\,\ref{sec:PTRTretrieval}, we evaluated the presence of individual molecular species in the atmosphere of \planet\ following the strategy outlined in \citep{benneke13}. 

The retrieval results are shown in Figure~\ref{fig:scarlet_cornerplot}.
Our analysis reveals that a Bayes factor of 2.6 in favor of the presence of water vapor in the atmosphere of \planet, which can be regarded as tentative evidence. No other molecular species is favored by the data. We also test for the presence of clouds by comparing the full retrieval model to a model that lacks clouds in the hypothesis space, but find no evidence in the observational data. The small differences between the evidence computed with SCARLET versus petitRADTRANS can be attributed in the difference in prior volume for the two analyses. We perform the final parameter estimation using the full retrieval model including the five molecules (H$_2$O, CH$_4$, CO$_2$, CO and N$_2$) and gray clouds. The best-fitting model  matches all data points within their 1-$\sigma$ uncertainties. A wide range of models is consistent with the data, in agreement with the results from petitRADTRANS.





\section{Cloud and Haze Models}
The retrieval analysis from the previous section showed that the muted water feature in the transmission spectrum is consistent with a low metallicity composition with high altitude condensates. To explore what condensate properties are plausible for \planet, we ran forward models with physically motivated cloud and haze opacity.

\subsection{Cloud Models}
Transmission spectra including the effect of clouds were calculated following the methodology of \cite{morley15, morley17}. First, 1D cloud-free model temperature profiles were calculated, assuming both radiative--convective and chemical equilibrium, using the approach described in detail in \citet{Mckay89, Marley99b, Saumon08, fortney08}. We calculate profiles for metallicities of [M/H]=0.0, 0.5, 1.0, 1.5, 2.0, and 2.5 (1, 3, 10, 30, 100, and 300$\times$ solar). The opacity database is described in detail in \citet{freedman08, freedman14}, with updated chemical equilibrium calculations and opacities as described in Marley et al. (in prep.). 

We include the condensation of Na$_2$S, KCl, and ZnS, which are expected to condense at the temperature of \planet\ ($T_\mathrm{eq} = 800$ Kelvin). We calculate cloud altitude and height along the cloud-free pressure--temperature profile; the cloud properties are calculated using the methods described in \citet{AM01, Morley12} for each metallicity, assuming a range of sedimentation efficiencies ($f_\mathrm{sed}$=2, 1, 0.5, and 0.1), a parameter which controls the cloud particle size and cloud height.  This model calculates the cloud optical depth, single-scattering albedo, and asymmetric parameter for each layer of the atmosphere. Example pressure--temperature profiles with cloud condensation curves are shown in Figure\,\ref{fig:PT}.

To calculate transmission spectra, we use the transmission spectrum model described in the appendix of \cite{morley17}. Gas opacity from  H$_2$ collisionally induced absorption,  CO$_2$,  H$_2$O,  CH$_4$,  CO,  NH$_3$,  PH$_3$,  H$_2$S,  Na,  K,  TiO,  VO,  and HCN is included. We calculate models for each metallicity and $f_\mathrm{sed}$ combination considered. 

\begin{figure*}
    \centering
    \includegraphics[scale = 0.75]{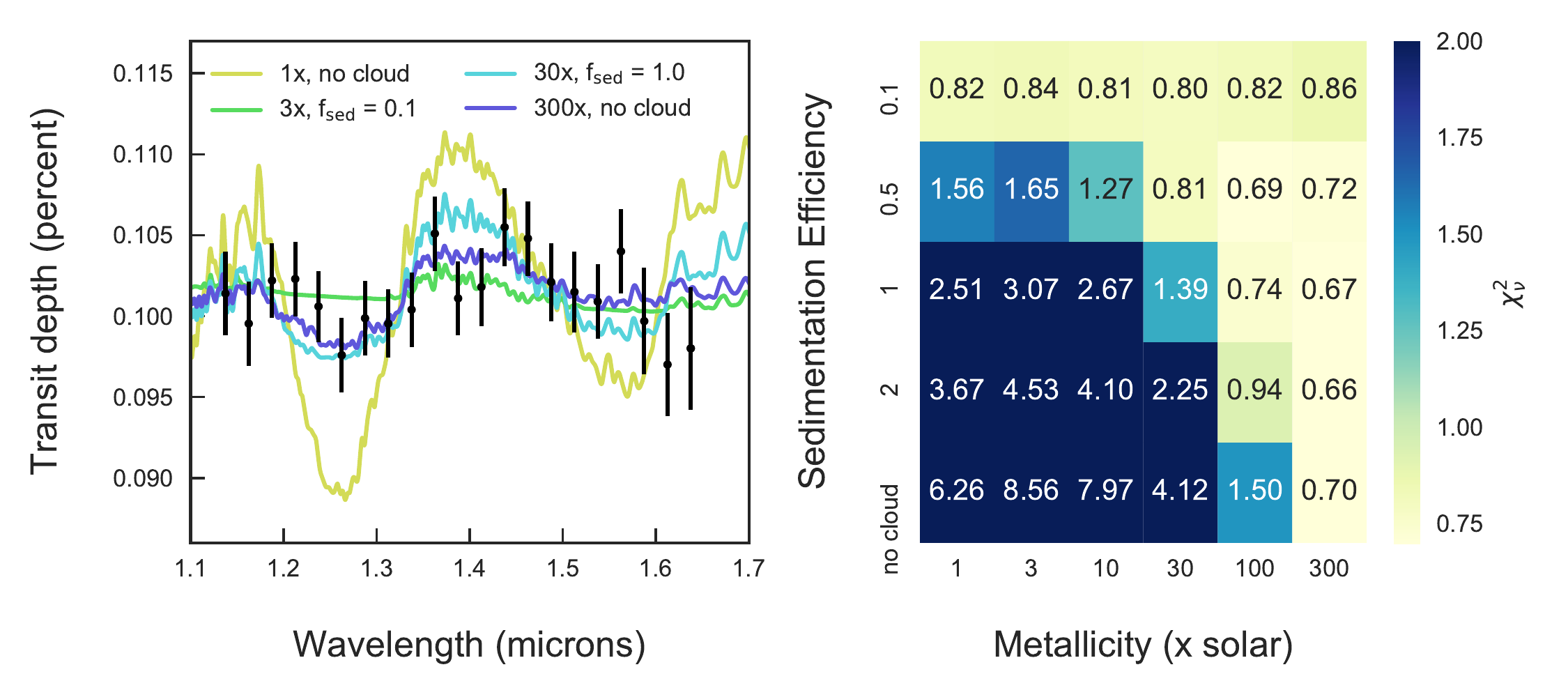}
    \caption{A sample of cloud forward models compared to the observed WFC3 spectrum (left) and a goodness-of-fit for the full grid of cloudy models (right). The grid cell shading indicates the reduced $\chi^2$ of the fit to the WFC3 data. The fit has 18 degrees of freedom (21 data points,  free parameters for metallicity, $f_\mathrm{sed}$, and reference radius).}
    \label{fig:cloudgrid}
\end{figure*}

Figure\,\ref{fig:cloudgrid} shows the goodness of fit for the cloudy model grid compared to the WFC3 transmission spectrum. The \Ktwo\ and \Spitzer\ data are not precise enough to significantly affect the goodness of fit. The best fits have small water absorption features with amplitude of around 30 ppm. The amplitude of features in the models is a trade-off between metallicity and cloud altitude: higher metallicity models tend to have a smaller scale height and thus smaller features. Lower sedimentation efficiency also decreases the feature amplitude. Small $f_\mathrm{sed}$ values loft cloud particles higher in the atmosphere, obscuring spectral features. As shown in Figure~\ref{fig:PT}, the cloud base is typically below the pressure level sensed by the observations, so $f_\mathrm{sed}$ values $\lesssim 0.5$ are required to loft the cloud into the observable atmosphere. For the \planet\ spectrum, the best fit models are high metallicity atmospheres ($100 - 300\times$ solar), or lower metallicity with high-altitude clouds ($f_\mathrm{sed} < 0.5$).  

\begin{figure}
    \centering
    \includegraphics[scale = 0.5]{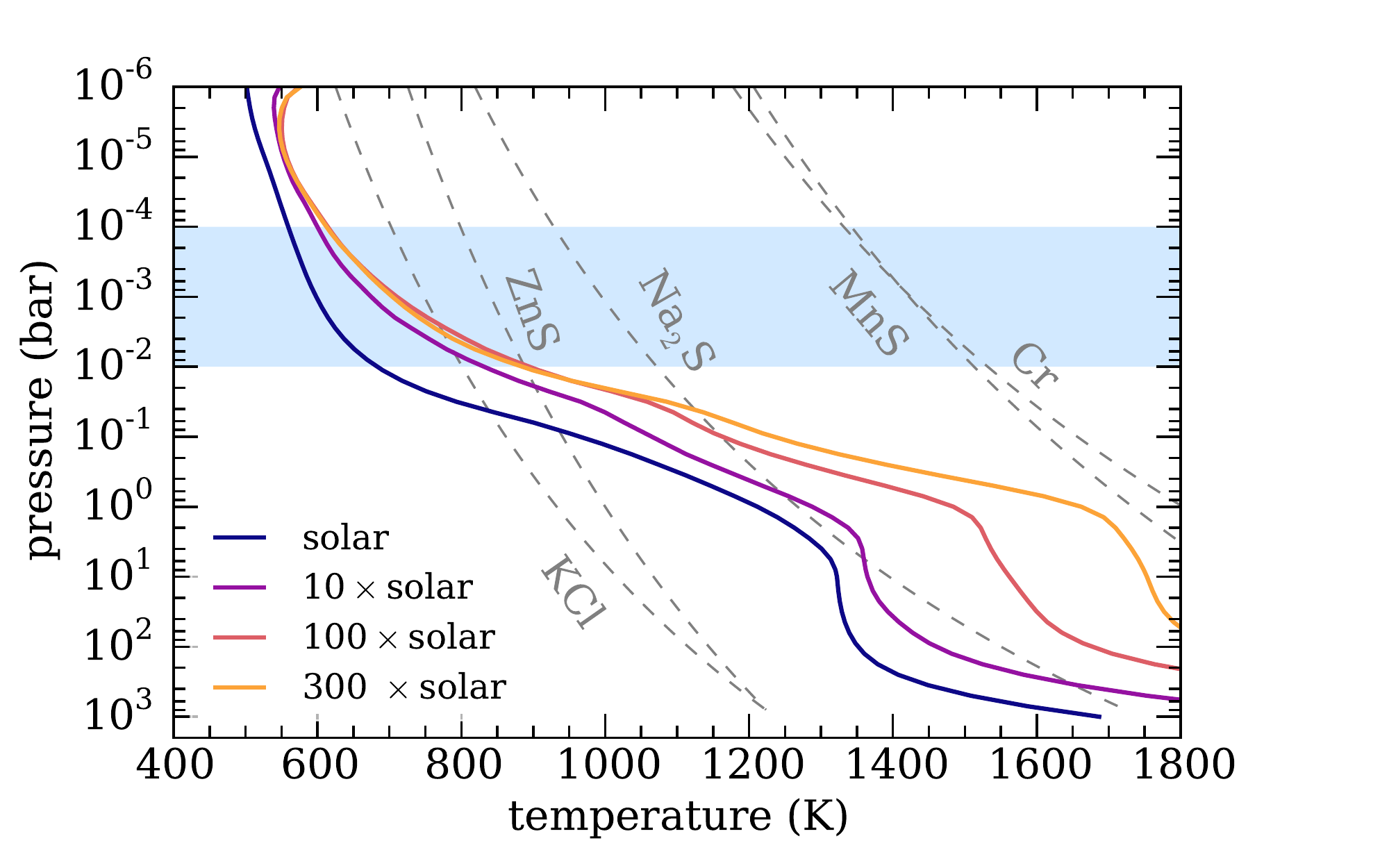}
    \caption{Pressure-temperature profiles (solid lines) for \planet\ compared to condensation curves for expected cloud species (dashed lines). The models assume are in radiative--convective and chemical equilibrium. The condensation curves are calculated for a 100$\times$ solar metallicity composition; for lower metallicities, the condensation curves shift left (by approximately 100 K per 1 dex metallicity). The shaded region marks the range of pressures sensed by transmission spectroscopy, assuming $100\times$ solar metallicity.}
    \label{fig:PT}
\end{figure}

\subsection{Haze Models}
We also calculate transmission spectra for hazy atmospheres using the photochemistry, microphysics, and transmission spectrum models of \cite{kawashima18} in the same way as \cite{kawashima19a} and \citet{kawashima19b}. We first perform photochemical simulations to derive the steady-state distribution of gaseous species. Then, since haze monomer production rate is uncertain for exoplanets, we assume a certain fraction of the sum of the photodissociation rates of the major hydrocarbons in our photochemical model, $\mathrm{CH_4}$, $\mathrm{HCN}$, and $\mathrm{C_2H_2}$, would result in haze monomer production. We call this fraction as haze formation efficiency $f_\mathrm{haze}$ following \citet{Morley13}.
With this assumption, we derive the steady-state distribution of haze particles by microphysical simulations.
Finally, we model transmission spectra of the atmospheres with the obtained profiles of haze particles and gaseous species.

For the temperature-pressure profile, we use the online-available\footnote{\url{http://cdsarc.u-strasbg.fr/viz-bin/qcat?J/A+A/562/A133}} analytical model of \citet{Parmentier14} assuming an internal temperature of 100~K and their correction factor of 0.25, which corresponds to the case where the irradiation is efficiently redistributed over the entire planetary surface. For the other input parameters, we use the default opacities \citep{Valencia13, Parmentier15} and Bond albedo. We include convection.
Solar elemental abundance ratios are taken from \cite{Lodders03}.
For the UV spectrum of HD~106315, we use that of the Sun from \cite{Segura03} because of its similar stellar type \citep[F5,][]{Houk99}. We assume a constant eddy diffusion coefficient of $10^7$~$\mathrm{cm}^2 \mathrm{s}^{-1}$ throughout the atmosphere for both photochemistry and microphysics calculations. We assume a monomer radius of 1~nm and an internal density of haze particles of 1~$\mathrm{g\ cm}^{-3}$.  The refractive index of haze is uncertain for exoplanets, so we consider two representative cases, tholin \citep{Khare84} and soot \citep{Hess98}.

\begin{figure*}
     \centering
     \includegraphics[scale = 0.25]{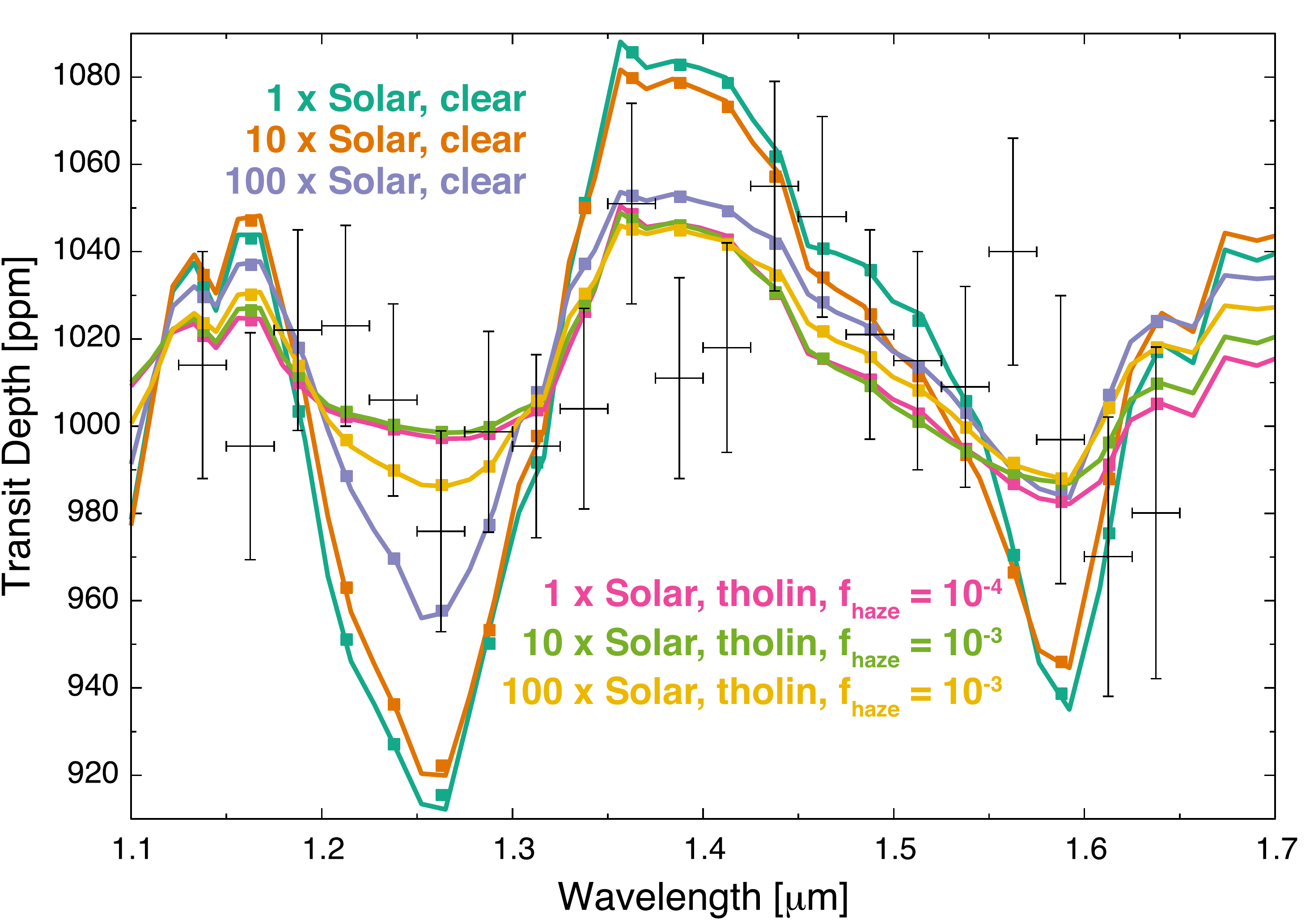}
     \includegraphics[scale = 0.28]{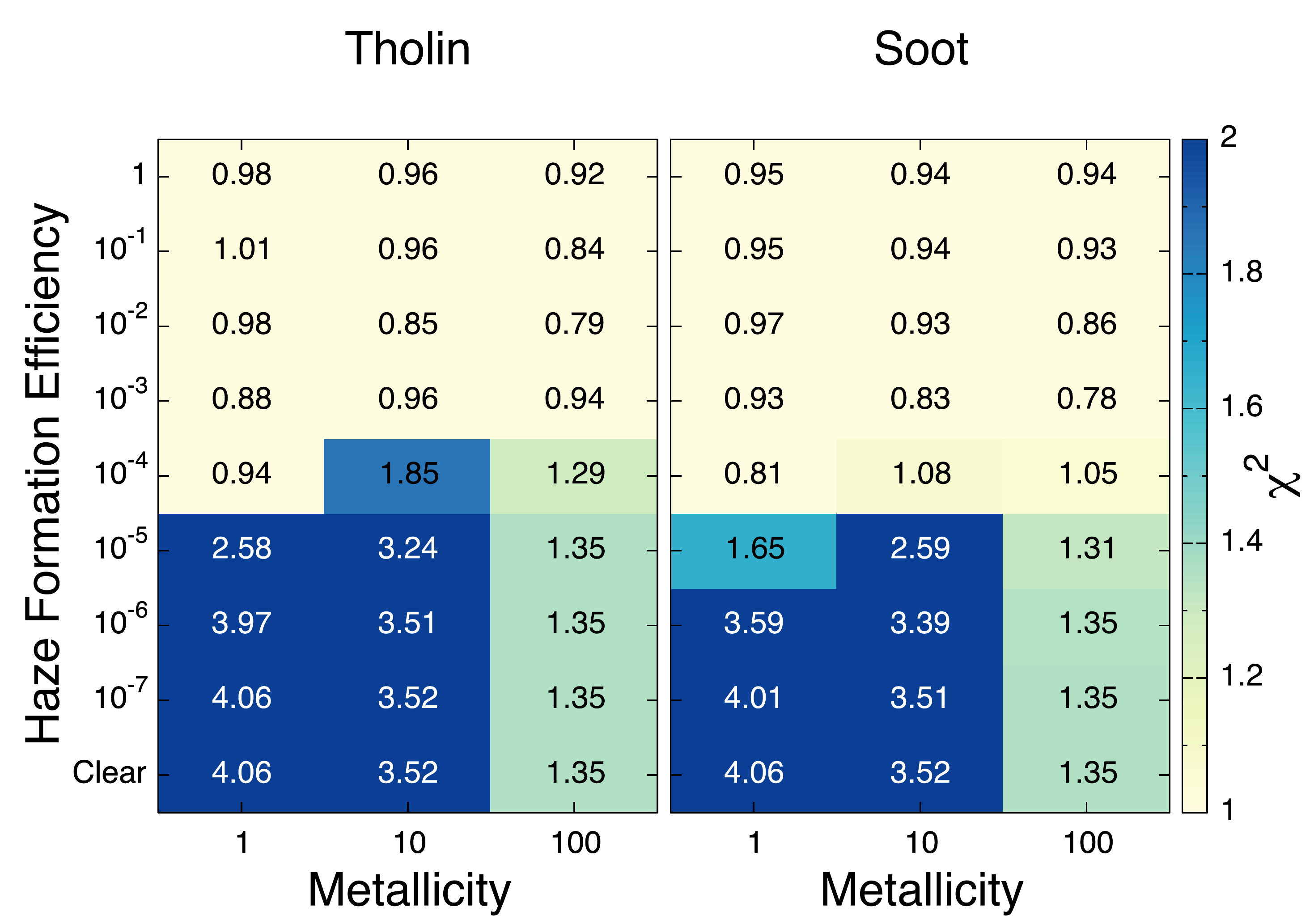}
    \caption{(Left)~Representative haze forward models compared to the measured WFC3 spectrum; models for the clear atmospheres of 1 (dark green line), 10 (orange line), and 100 (purple line) $\times$ Solar metallicity and those for the hazy (tholin) atmospheres of 1 (pink line), 10 (light green line), and 100 (yellow line) $\times$ Solar metallicity. The haze formation efficiency that fits the observed data well is chosen. (Right)~Goodness-of-fit for the full grid of tholin and soot models. The grid cell shading indicates the reduced $\chi^2$ of the fit to the WFC3 data. The fit has 18 degrees of freedom (21 data points, free parameters for metallicity, $f_\mathrm{haze}$, and reference radius). 
    }
   \label{fig:haze}
\end{figure*}

We calculate spectra for 1, 10, and 100 $\times$  solar metallicity atmospheres with a range of $f_\mathrm{haze}$ from $10^{-7}$ to $1$ in 1 dex increments. The integrated monomer production rate for $f_\mathrm{haze} = 1$ (the sum of the photodissociation rates of $\mathrm{CH_4}$, $\mathrm{HCN}$, and $\mathrm{C_2H_2}$) becomes smaller with increasing metallicity; $1.71 \times 10^{-10}$, $1.20 \times 10^{-10}$, and $5.51 \times 10^{-11}$ $\mathrm{g}$~$\mathrm{cm^2}$~$\mathrm{s}^{-1}$ for 1, 10, and 100$\times$ the solar metallicity atmospheres, respectively. 

For the calculation of transmission spectra, we treat the reference radius at 10~bar pressure level as a parameter. We find the appropriate value with a grid of 0.1\% of the observed transit radius 
which yields the minimum reduced ${\chi}^2$ with 18 degrees of freedom (21 data points minus 3 free parameters of metallicity, $f_\mathrm{haze}$, and reference radius), for each case. We account for the transmission curve of the WFC3 G141 grism
from the SVO Filter Profile Service\footnote{\url{http://svo2.cab.inta-csic.es/theory/fps/}} \citep{2012ivoa.rept.1015R, Rodrigo2013}.

The left panel of Figure\,\ref{fig:haze} shows several representative models compared to the measured WFC3 spectrum; models for clear atmospheres of three different metallicities, as well as hazy (tholin) atmospheres with haze formation efficiency tuned to fit the WFC3 data well. The error bars for the \Ktwo\ and \Spitzer\ points are large, and therefore have a negligible effect on the goodness of fit. The right panel of Figure\,\ref{fig:haze} shows the goodness of fit for the model grids. 
We find that modest haze formation efficiencies of $10^{-3} - 10^{-4}$ generally match the observed spectra for all the three different metallicities, for both tholins and soots. This is because smaller scale height due to increasing metallicity can be compensated out by smaller fiducial monomer production rate.
Overall, these haze production efficiencies are orders of magnitude lower than the extreme values required to reproduce the featureless spectrum of GJ~1214b \citep{morley15, kawashima19a}. As noted above, the NUV irradiation of \planet\ is likely higher than that of GJ~1214b, so more haze precursors are present and lower haze production efficiency is needed.

\section{Interior structure models}
\begin{figure}
    \centering
    \includegraphics[scale = 0.5]{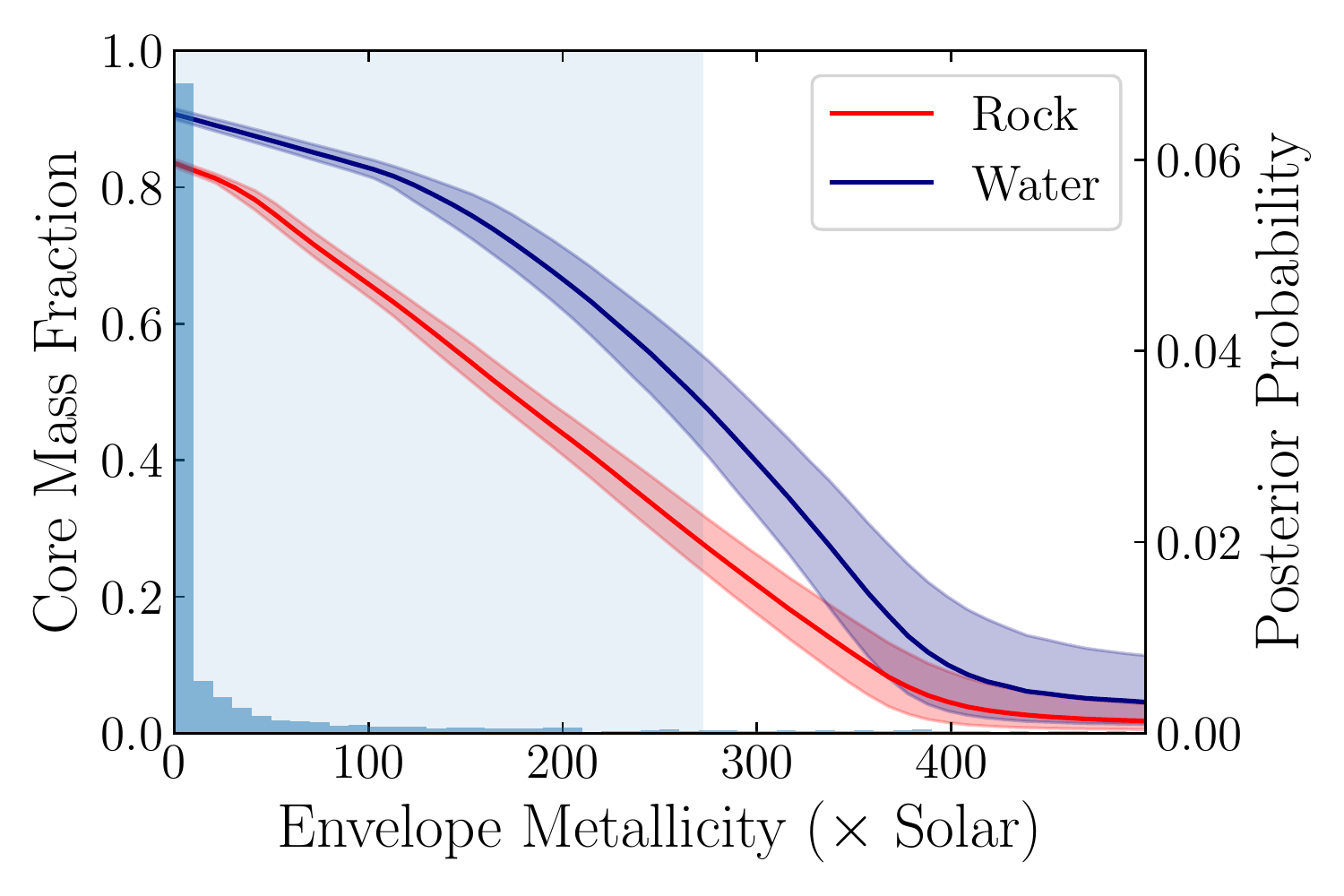}
    \caption{Core fraction versus envelope metallicity from interior structure modelling for a rocky core (red line with $1\sigma$ uncertainty shaded) and a water core (blue line with $1\sigma$ uncertainty).   The retrieval results for the envelope metallicity are over-plotted as a histogram, with the $1\sigma$ credible interval indicated by the blue shaded region.}
    \label{fig:interior}
\end{figure}

Comparison between interior structure and envelope metallicity can provide additional constraints on the bulk composition of the planet \citep{kreidberg18b, thorngren19}. For example, given knowledge of the envelope metallicity, it is possible to put limits on the core mass, that otherwise suffers from large degeneracy for planets in the $2 -5\,R_\oplus$ radius range \citep{rogers10}. We evaluate the internal structure of \planet\ with a model consisting of an inner core and a H/He outer envelope enriched with some various amounts of water and rock (in a 50-50 ratio), using the methods described by \citep{thorngren16}.  We explored two limiting cases for the core composition: one is composed entirely of isothermal rock with radioactive heating, and the other is composed of convective water.  Using the observed mass (with error), radius (with error), age (with error), and flux (ignoring error), we retrieved the core mass over a range of envelope metallicities. Our results are shown in Figure~\ref{fig:interior}.

In the absence of any information about the envelope composition, the core mass fraction for \planet\ could range anywhere from 0 to 1. The higher the envelope metallicity, the lower the core mass fraction required to explain the observed mass, radius, and age of the planet. To help break  this degeneracy, we compared the results from the atmospheric retrieval with the interior structure model (using water abundance as a proxy for envelope metallicity). The retrieval results are shown alongside the interior structure model in Figure~\ref{fig:interior}.  Using the retrieved abundance of H$_2$O from petitRADTRANS as a proxy for the envelope metallicity ($5\times10^{-4} - 290\times$ solar at $1\sigma$ confidence), we estimate that the core mass fraction is greater than 30\% regardless of the core composition (rock or ice). These conclusions generally resemble our understanding of the bulk composition of Uranus and Neptune, which are expected to have a core mass of 80 - 90\% \citep{hubbard91, fortney10, nettelmann13}. Follow-up atmosphere characterization with higher precision and broader wavelength coverage can further constrain the envelope metallicity and core mass of \planet.

\section{Discussion and Conclusions}
The number of small exoplanets with precise transmission spectra is growing, and already the population shows diversity in atmospheric properties. Some appear to have envelope metallicities below that of Neptune \citep[e.g. HAT-P-26b;][]{wakeford17}, whereas others require higher metallicity \citep[GJ~436b;][]{morley17}. Some planets are blanketed with thick high altitude clouds or haze \citep[particularly GJ~1214b;][]{kreidberg14a}, while others have deeper condensates or even cloud-free atmospheres \citep{benneke19, madhusudhan20}.  This diversity is expected from theoretical models. For example, planet population synthesis predicts a wide range of envelope enrichment for sub-Neptunes \citep[e.g][]{fortney13}. Similarly, cloud and haze models indicate that condensate properties may vary widely across different planets. Condensate formation depends on many different atmospheric properties (e.g. temperature, metallicity, UV irradiation and vertical mixing) so there is no one-size-fits-all model to predict whether an atmosphere is cloudy or clear at the pressure levels sensed by transmission spectroscopy \citep{morley15,gao18, he18, horst18, kawashima19a, ohno20}.  

Where does \planet\ fit into this diverse picture? The small amplitude of spectral features is consistent with other sub-Neptunes, which all have feature amplitudes attenuated relative to expectations for solar composition, cloud-free atmospheres \citep{crossfield17}. Intriguingly, the amplitude of spectral features ($0.8\pm0.0.04$ H$_2$/He scale heights) agrees well with the demographic trend noted in \cite{crossfield17, roberts20}, that shows an increase in the amplitude of the WFC3 water feature with planet equilibrium temperature. This is a somewhat surprising finding, because there are many factors (noted above) that affect the observed spectral feature amplitude for planets in this population. A further demographic study of water absorption in sub-Neptunes will be explored in a follow-up paper.

The tentative water detection for \planet\ is consistent with a wide range of abundances ($3\times10^{-4} - 290\times$ solar at 1$\sigma$ confidence), and is most comparable to that estimated for HAT-P-11b and K2-18b \citep{fraine14, benneke19, chachan19}. Low metallicities ($<50\times$ solar), akin to those GJ~3470b, HAT-P-26b, and WASP-107b \citep{benneke19a, wakeford17, kreidberg18} are possible for \planet, provided it has some condensates in its atmosphere that truncate the amplitude of the water feature. The condensate properties are modest relative to extremes like GJ~1214b, which has a featureless spectrum requiring either very low sedimentation efficiency clouds and high atmospheric metallicity ($f_\mathrm{sed} \le 0.1$ and $1000\times$ solar composition), or very efficient photochemical haze production ($\gtrsim10\%$ efficiency) for a $50\times$ solar metallicity composition \citep{kreidberg14a, morley15}.  For comparison, the transmission spectrum of \planet\ is fit well with $f_\mathrm{sed} \le 0.5$ or haze production efficiencies of $10^{-3} - 10^{-4}$.

The tentative detection of a small water feature in \planet\ makes it an intriguing candidate for follow-up observations to further characterize its atmosphere.  Infrared observations are a particularly promising avenue --- spectroscopy in the $4 - 5\mu$m range is sensitive to absorption from CO$_2$, a prominent feature expected in high metallicity atmospheres \citep{moses13}. If the atmosphere has lower metallicity but is cloudy/hazy, infrared observations are also promising because the condensates may have lower opacity at longer wavelengths \citep[e.g GJ 3470b;][]{benneke19}.   Future transmission spectroscopy observations with \JWST\ could potentially distinguish between these possibilities \citep{greene16}, and confirm whether \planet\ does indeed have a Neptune-like core mass and envelope composition. If it does, that will provide new incentive for formation models to produce ice giants on a wide range of orbits from 0.15 AU (that of \planet) to 30 AU (that of Neptune).

\acknowledgements
Support for HST program GO-15333 was provided by
NASA through a grant from the Space Telescope Science Institute, which is operated by the Association of Universities for Research in Astronomy, Inc., under NASA contract NAS 5-26555. This work is based [in part] on observations made with the Spitzer Space Telescope, which was operated by the Jet Propulsion Laboratory, California Institute of Technology under a contract with NASA. This research has made use of the SIMBAD database, operated at CDS, Strasbourg, France. Some of the data presented in this paper were obtained from the Mikulski Archive for Space Telescopes (MAST). STScI is operated by the Association of Universities for Research in Astronomy, Inc., under NASA contract NAS5-26555. Support for MAST for non-HST data is provided by the NASA Office of Space Science via grant NNX13AC07G and by other grants and contracts.  This research made use of matplotlib, a Python library for publication quality graphics \citep{Hunter:2007} This research made use of SciPy \citep{Virtanen_2020} This research made use of NumPy \citep{van2011numpy}. P.M. acknowledges support from the European Research Council under the European Union's Horizon 2020 research and innovation program under grant agreement No. 832428. Y.K. acknowledges support from the European Union’s Horizon 2020 Research and Innovation Programme under Grant Agreement 776403. L.K. acknowledge M.R.~Line for illuminating discussions.

\bibliographystyle{aasjournal}



\end{document}